\documentclass[twocolumn]{aastex63}

\usepackage{bm}
\usepackage{url}
\usepackage{amsmath}
\usepackage{color}
\usepackage{graphicx}
\usepackage{threeparttable}

\usepackage{ulem}

\received{}
\revised{}
\accepted{}

\shorttitle{Inner solar system formation}
\shortauthors{Ueda et al.}

\begin{document}

\title{
Early Initiation of Inner Solar System Formation at Dead-Zone Inner Edge
}

\correspondingauthor{Takahiro Ueda}
\email{takahiro.ueda@nao.ac.jp}

\author[0000-0003-4902-222X]{Takahiro Ueda}
\affil{National Astronomical Observatory of Japan, Osawa 2-21-1, Mitaka, Tokyo 181-8588, Japan
}

\author[0000-0002-8300-7990]{Masahiro Ogihara}
\affil{National Astronomical Observatory of Japan, Osawa 2-21-1, Mitaka, Tokyo 181-8588, Japan
}
\affil{Earth-Life Science Institute, Tokyo Institute of Technology, Meguro-ku, Tokyo 152-8550, Japan}

\author[0000-0002-5486-7828]{Eiichiro Kokubo}
\affil{National Astronomical Observatory of Japan, Osawa 2-21-1, Mitaka, Tokyo 181-8588, Japan
}

\author[0000-0002-1886-0880]{Satoshi Okuzumi}
\affil{Department of Earth and Planetary Sciences, Tokyo Institute of Technology, Meguro, Tokyo, 152-8551, Japan
}



\begin{abstract}
The inner solar system possesses a unique orbital structure in which there are no planets inside the Mercury orbit and the mass is concentrated around the Venus and Earth orbits.
The origins of these features still remain unclear.
We propose a novel concept that the building blocks of the inner solar system formed at the dead-zone inner edge in the early phase of the protosolar disk evolution, where the disk is effectively heated by the disk accretion.
First, we compute the dust evolution in a gas disk with a dead zone and obtain the spatial distribution of rocky planetesimals.
The disk is allowed to evolve both by a viscous diffusion and magnetically-driven winds.
We find that the rocky planetesimals are formed in concentrations around $\sim$ 1 au with a total mass comparable to the mass of the current inner solar system in the early phase of the disk evolution within $\lesssim0.1$ Myr.
Based on the planetesimal distribution and the gas disk structure, we subsequently perform \textit{N}-body simulations of protoplanets to investigate the dynamical configuration of the planetary system.
We find that the protoplanets can grow into planets without significant orbital migration because of the rapid clearing of the inner disk by the magnetically-driven disk winds.
Our model can explain the origins of the orbital structure of the inner solar system.
Several other features such as the rocky composition can also be explained by the early formation of rocky planetesimals.
\end{abstract}



\section{Introduction}
The solar system is the most familiar planetary system and has been studied well for the past decades \citep[e.g.,][]{Hayashi1981,2001Icar..152..205C,Raymond+09,2010Icar..207..517M,2014RSPTA.37230174J,Lichtenberg+21}.
However, formation of the solar system is still the most interesting subject.
The inner solar system possesses a unique dynamical configuration: large planets (Venus and Earth) sandwiched by two small planets (Mercury and Mars) and absence of planets inside the orbit of Mercury.
This configuration suggests that the building blocks of the inner solar system planets were not distributed uniformly but locally (\citealt{Raymond+09,Hansen2009}).
\citet{Hansen2009} proposed a scenario that the inner solar system planets were formed from a narrow planetesimal annulus ranging from $0.7$ to $1~{\rm au}$ with a total mass of $2M_{\oplus}$. 
This model can explain the configuration of the inner solar system planets.

Although several mechanisms have been proposed for the formation of a narrow annulus of rocky planetesimals (e.g., \citealt{Drazkowska2016,2018A&A...612L...5O}), dust pileup at the dead-zone inner edge is one of the most preferred models for rocky planetesimal formation (e.g., \citealt{Kretke+09,CT14,Ueda2019}).
At the dead-zone inner edge, the gas temperature reaches $\sim$ 800--1000~K (e.g., \citealt{Desch2015}; \citealt{jankovic21}), above which thermal ionization of the gas is sufficiently effective to activate magnetorotational instability (MRI) \citep{Gammie1996}.
Across the dead-zone inner edge, the turbulent viscosity induced by the MRI steeply decreases from inside out, resulting in a local maximum in the radial profile of the gas pressure (e.g., \citealt{Dzyurkevich2010}; \citealt{Flock2017}), which traps solid particles \citep{Whipple1972, Adachi1976}, and hence, leads to the formation of rocky planetesimals by streaming instability \citep{Youdin2005} and/or gravitational instability \citep{CMF1981}.

Even if planetesimals form with the distribution proposed by \citet{Hansen2009}, their subsequent evolution depends on the evolution of the gas disk because protoplanets/planets interact with the gas disk gravitationally.
One of the major difficulties is the so-called Type-I migration in which an Earth-sized planet at 1 au moves radially inward within a timescale of $\lesssim$ 1 Myr (e.g., \citealt{Ward1997,TTW2002}).
\citet{Ogihara+15,Ogihara2018} showed that Type-I migration can be suppressed and Earth-sized planets can survive at approximately 1 au if the magnetically-driven disk winds dissipate the inner disk gas. 

In this letter, we propose a new concept that the formation of terrestrial planets starts at the dead-zone inner edge in the early phase of the disk evolution.
The dead-zone inner edge is located at $\sim0.1$ au for a typical passive T-Tauri disk (e.g., \citealt{Ueda2017}); however, it can be located beyond 1 au in the early phase of the disk evolution during which the accretion heating is effective. 
We demonstrate that rocky planetesimals can form around the current Earth orbit and grow into planets without significant orbital migration in the presence of the magnetically-driven disk winds.
Our calculation is divided into two parts.
In Section \ref{sec:dust2plts}, we show the setup and results of the dust growth simulations.
Subsequently, \textit{N}-body simulations based on the obtained planetesimal distribution are presented in Section \ref{sec:plts2planets}.
The discussion and conclusions are presented in Sections \ref{sec:Discussion} and \ref{sec:Conclusion}, respectively.

\section{From dust to planetesimals} \label{sec:dust2plts}
In this section, we describe our dust-growth simulation method and subsequently present the time evolution of the protosolar disk as well as the obtained planetesimal distribution.

\subsection{Simulation setup: dust and gas disk evolution}\label{sec:dust2plts-setup}
We calculate dust evolution in a gas disk around a Sun-like star evolved with the magnetically-driven disk winds as well as viscous diffusion.
The method for the dust-growth simulation is the same as that used by \citet{Ueda2019}. 
In this method, dust evolution in the gas disk is calculated based on a single-size approach, in which we trace the evolution of the maximum dust size in each radial grid \citep{Sato2016}.
The dust grains are evolved by direct sticking, fragmentation, radial advection, and turbulent diffusion.
The critical fragmentation velocities of silicate and icy grains are set as 3 and 10 ${\rm m~s^{-1}}$, respectively.
The silicate and water-ice components are assumed to sublimate when the temperatures reach $150$ and $1500~{\rm K}$, respectively.
Just after the ice sublimation, silicate grains are assumed to have a size that is determined by the turbulence-induced fragmentation.

We employ a planetesimal formation algorithm described in \citet{Ueda2019} (see also \citealt{Drazkowska2016}); if the mid-plane dust-to-gas mass ratio exceeds unity, dust is converted into planetesimals with a timescale of $\zeta^{-1} T_{\rm K}$ where we adopt $\zeta=10^{-4}$ as a fiducial case.

We adopt the weak disk-wind model proposed by (\citealt{Suzuki2016}; see also \citealt{Kunitomo+20,Taki+21}) as the gas disk model.
We also implement the effect of a dead-zone in the gas disk.
MRI is assumed to occur if the disk mid-plane temperature is above $T_{\rm MRI}=800~{\rm K}$ or the gas surface density is below 20 ${\rm g~cm^{-2}}$.
The former represents the effect of the thermal ionization of the disk gas and sets the dead-zone inner edge.
The latter represents the effect of the non-thermal ionization induced by cosmic rays. 
The disk mid-plane is heated by the stellar irradiation and disk accretion.
For the dust opacity, we adopt the opacity model given by \citet{Kunitomo+20}, where the opacity is fixed to $4.5~{\rm cm^{2}~g^{-1}}$ for the rocky region.
We adopt a turbulence strength of $\alpha_{\rm MRI}=2\times10^{-2}$ for the MRI-active region and $\alpha_{\rm DEAD}=2.3\times10^{-4}$ for the MRI-dead region as a fiducial model.
As the initial condition, we assume a compact disk; the gas surface density follows a radial power-law index of 1.5 and a exponential tail with a cut-off radius of 15 au. 
The initial disk mass and the dust-to-gas mass ratio are set as $0.1M_{\odot}$ and 0.01, respectively.

\subsection{Disk evolution and planetesimal distribution}\label{sec:dust2plts-results}

\begin{figure}[ht]
\begin{center}
\includegraphics[scale=0.55]{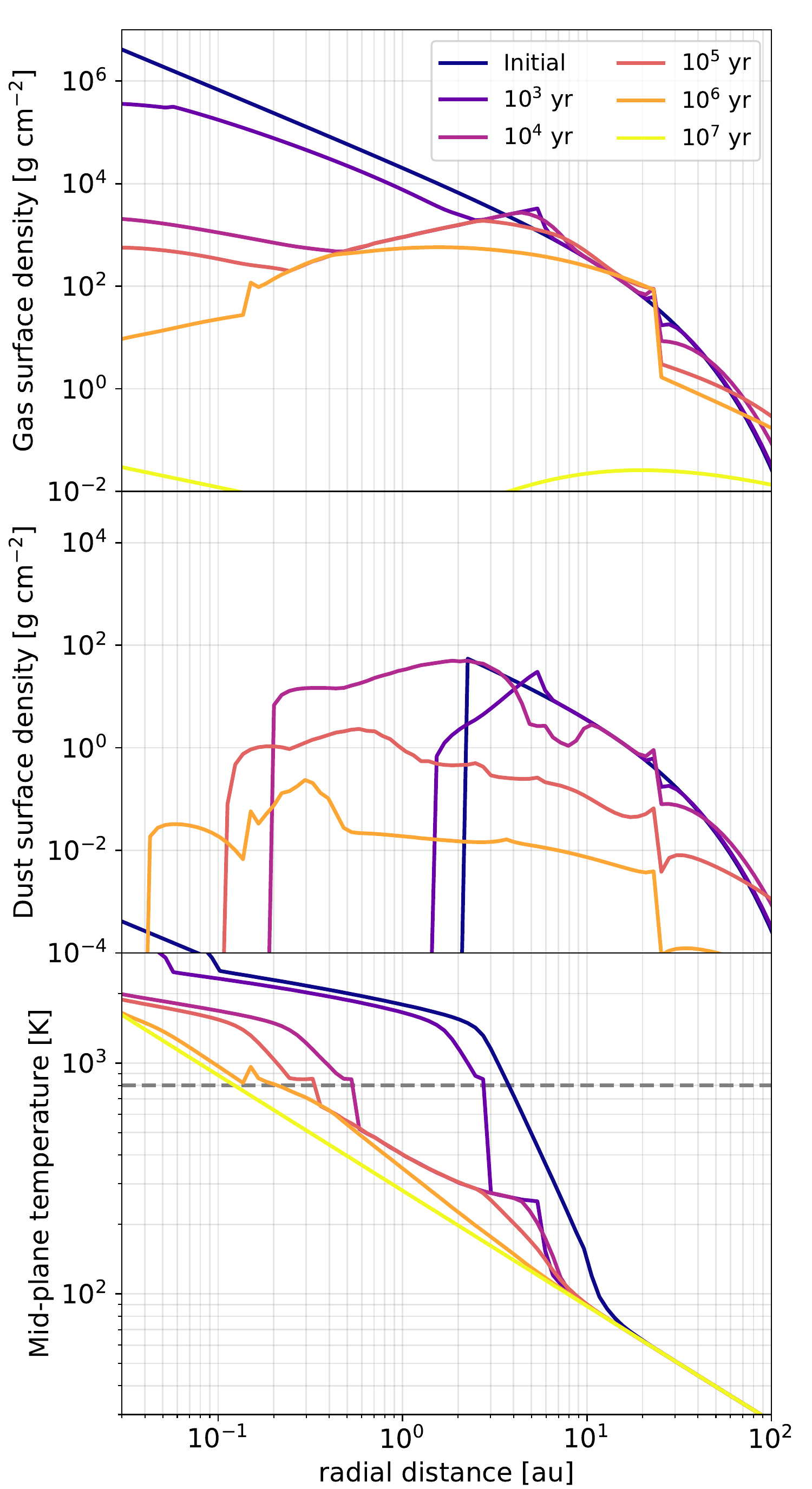}
\caption{
Time evolution of gas surface density (top), dust surface density (middle), and mid-plane temperature (bottom).
}
\label{fig:disk}
\end{center}
\end{figure}

Figure \ref{fig:disk} shows the time evolution of the gas surface density, dust surface density, and mid-plane temperature of the disk.
In the beginning of the disk evolution, the gas surface density evolution is governed by the turbulent viscosity.
The inner region of the disk is effectively heated by the disk accretion, and the dead-zone inner edge is located at $r \sim$ 4 au, where $r$ represents the radial distance from the star.
Inside the dead-zone inner edge, the disk rapidly dissipates owing to the high viscosity, whereas the region outside the dead-zone inner edge is almost stationary until $t\sim$ 1 Myr.
The outermost region, $r\gtrsim25$ au, also diffuses rapidly owing to the high viscosity because of the high ionization degree induced by the cosmic rays.
As the gas surface density decreases, the mid-plane temperature also decreases, and hence, the dead-zone inner edge moves inward with time.
In the early phase ($t\lesssim10^{-5}$), the gas surface density at $\sim$ 0.2-4 au is shaped by the migrating dead-zone inner edge. 
Since the dead-zone inner edge migrates, the gas surface density gradient is more shallower than that obtained from a quasi-static model (e.g., \citealt{Ueda2019}). 
At $t\sim$ 10 kyr, rocky grains start to pile up at $\sim$ 2 au.
With the given turbulence strength ($\alpha_{\rm DEAD}=2.3\times10^{-4}$), rocky grains do not pile up strongly; however, they accumulate marginally just outside the dead-zone inner edge.
Following the radial motion of the dead-zone inner edge, dust accumulation also moves inward: $\sim$ 2, 0.7, and 0.3 au at $10^{4}, 10^{5}$, and $10^{6}$ yr, respectively.

\begin{figure}[ht]
\begin{center}
\includegraphics[scale=0.4]{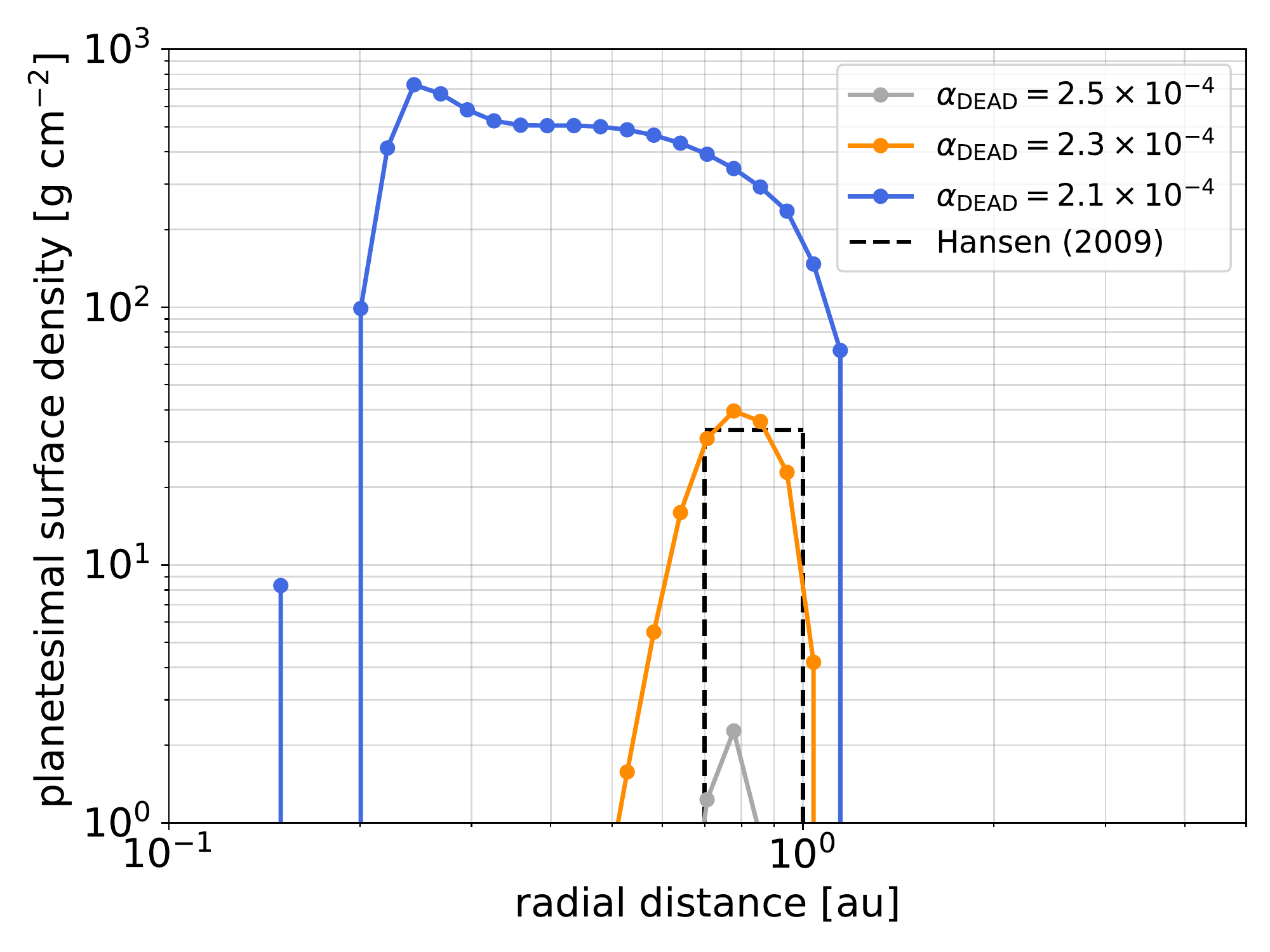}
\caption{
Obtained planetesimal surface density profile with $\alpha_{\rm DEAD}=2.3\times10^{-4}$ (orange).
The black dashed line corresponds to planetesimal distribution proposed by \citet{Hansen2009}.
For reference, results of simulations with $\alpha_{\rm DEAD}=2.1\times10^{-4}$ (blue) and $2.5\times10^{-4}$ (gray) are also shown.
}
\label{fig:plts}
\end{center}
\end{figure}

Figure \ref{fig:plts} shows the planetesimal distribution obtained from the simulation of the dust evolution.
For reference, we also show the results with $\alpha_{\rm DEAD}=2.1\times10^{-4}$ and $2.5\times10^{-4}$.
We clearly see that rocky planetesimals form around the solar system terrestrial planet region.
For $\alpha_{\rm DEAD}=2.3\times10^{-4}$, the surface density profile of the formed planetesimals is comparable to that proposed by \citet{Hansen2009}.
The inner edge of the planetesimal belt is at $r \simeq$ 0.6 au, and no planetesimals form inside it.
If the turbulence is stronger ($\alpha_{\rm DEAD}=2.5\times10^{-4}$), the dust-pileup is suppressed because of the stronger turbulent mixing, and hence, the total mass of the planetesimals is smaller.
In contrast, if the turbulence is weaker ($\alpha_{\rm DEAD}=2.1\times10^{-4}$), rocky grains pileup efficiently, and hence, significant amount of planetesimals form.

The outer edge of the planetesimal belt is determined by the location of the dead-zone inner edge when the pileup starts.
In our model, the disk is initially massive and compact (mass of $0.1M_{\odot}$ and radius of 15 au), and hence, the dead zone can be located beyond 1 au.
The inner edge of the planetesimal belt is determined by the balance between the radial mass flux of the drifting pebbles and the turbulent mixing.
Because the dust mass flux is a decreasing function of time \citep{LJ14}, the turbulent mixing dominates the accumulation of dust at some point.
Because we select the turbulence strength to be such that the dust pileup is marginally ($\alpha_{\rm DEAD}=2.3\times10^{-4}$), the planetesimals concentrate at $r \sim$ 0.6--1 au, which is preferable for the formation of solar system terrestrial planets.
If the turbulence strength is sufficiently weak, planetesimals continuously form at the moving dead-zone inner edge, and the planetesimal belt reaches $r \sim$ 0.1 au.
This broad and massive planetesimal belt will be unsuitable for the inner solar system; however, it will be preferable for the formation of close-in super-Earth systems.

The strong dependence of the planetesimal distribution on the turbulence strength indicates that there are two regimes in the planet formation at the dead-zone inner edge.
In the weak turbulence regime ($\alpha_{\rm DEAD}\lesssim2.1\times10^{-4}$), rocky planetesimals efficiently form in the inner region of the disk, which can account for super-Earth systems.
In the strong turbulence regime ($\alpha_{\rm DEAD}\gtrsim2.5\times10^{-4}$), no planetesimals/planets form in the inner region of the disk.
The inner solar system lies in between these two regimes.

\subsection{Parameter dependence}\label{sec:param}
So far, we focus on our fiducial model which is the best for inner solar system formation in our calculations. 
However, it would be worth to describe the parameter dependence on the planetesimal distribution.
Table \ref{table:0} summarizes parameters and planetesimal distribution (total mass $M_{\rm plts}$ and peak position in planetesimal surface density $r_{\rm peak}$) at $t=10^{5}~{\rm yr}$ obtained in different models.
\begin{table}[ht]
\begin{center}
\caption{Parameter dependence on planetesimal distribution}
\label{table:0}
\begin{tabular}{cccccccc}
  \hline  
ID&$\alpha_{\rm DEAD}$ & $v_{\rm f}$ & $r_{\rm d}$ & $\zeta$   & $M_{\rm plts}$  & $r_{\rm peak}$ \\
  &                    & [m/s]       & [au]        &           & [$M_{\oplus}$]  & [au]     \\ \hline  \hline
0&$2.3\times10^{-4}$  & 3           & 15          & $10^{-4}$ & 2.2             & 0.78      \\
1&$2.1\times10^{-4}$  & 3           & 15          & $10^{-4}$ & 48              & 0.24      \\
2&$2.5\times10^{-4}$  & 3           & 15          & $10^{-4}$ & 0.06            & 0.78      \\
3&$3.0\times10^{-4}$  & 3           & 15          & $10^{-4}$ & 0               & N/A      \\\hline
4&$2.3\times10^{-4}$  & 1           & 15          & $10^{-4}$ & 0               & N/A      \\
5&$2.3\times10^{-4}$  & 10          & 15          & $10^{-4}$ & 78              & 0.64      \\\hline
6&$2.3\times10^{-4}$  & 3           & 30          & $10^{-4}$ & 2.8             & 0.71      \\
7&$2.3\times10^{-4}$  & 3           & 100         & $10^{-4}$ & 13.8            & 0.53      \\\hline
8&$2.3\times10^{-4}$  & 3           & 15          & $10^{-3}$ & 6.7             & 0.78      \\
9&$2.3\times10^{-4}$  & 3           & 15          & $10^{-5}$ & 0.28            & 0.78      \\
  \hline  
\end{tabular}
\end{center}
\end{table}

The total planetesimal mass is also sensitive to the critical fragmentation velocity $v_{\rm f}$.
If $v_{\rm f}=1~{\rm m~s^{-1}}$, no planetesimals form, while planetesimals with total mass of 78$M_{\rm \oplus}$ form when $v_{\rm f}=10~{\rm m~s^{-1}}$, in our model.
This is because lower $v_{\rm f}$ makes dust smaller, leading to inefficient dust-trapping.
The total planetesimal mass is an increasing function of the planetesimal formation efficiency $\zeta$; $M_{\rm plts}=0.28, 2.2$ and 6.7$M_{\oplus}$ for $\zeta=10^{-5},10^{-4}$ and $10^{-3}$.
It would be worth to emphasize that the total planetesimal mass can decrease (increase) to $2M_{\oplus}$ even with $\zeta=10^{-3}$ by slightly increasing (decreasing) $\alpha_{\rm DEAD}$.
On the other hand, the location where planetesimal form is mainly determined by the gas surface density and hence the disk size for a given disk mass ($0.1M_{\odot}$ in our model).
The gas disk needs to have a initial radius of $<30$ au to form a narrow planetesimal belt at current terrestrial planet region.

\section{From protoplanets to planets} \label{sec:plts2planets}
In this section, we perform \textit{N}-body simulations of planet formation to investigate the subsequent evolution of formed planetesimals.

\subsection{Simulation setup: \textit{N}-body simulations}\label{sec:plts2planets-setup}
As the initial condition, protoplanets with masses of $0.01 \,M_{\oplus}$ and eccentricities/inclinations of $\simeq 0.01$ are distributed with a surface density profile that is the same as the planetesimal distribution presented in Section \ref{sec:dust2plts}.
We use the gas disk structure shown in Figure\,\ref{fig:disk} and start \textit{N}-body simulations at time of 30 kyr. Although not discussed in this letter, additional simulations are performed in which \textit{N}-body simulations are initiated at time of 1 Myr with the same distibution of protoplanet with our fiducial model, which reflects the time elapse for planetesimals to grow into protoplanets.
We confirmed that this does not influence our conclusions.
In the calculation, we compute the gravitational interaction between protoplanets orbiting around the Sun and assume a perfect accretion when they collide.
We also consider planet-disk interaction.

\subsection{Evolution of protoplanets}\label{sec:plts2planets-results}
\begin{figure}[ht]
\begin{center}
\includegraphics[scale=0.57]{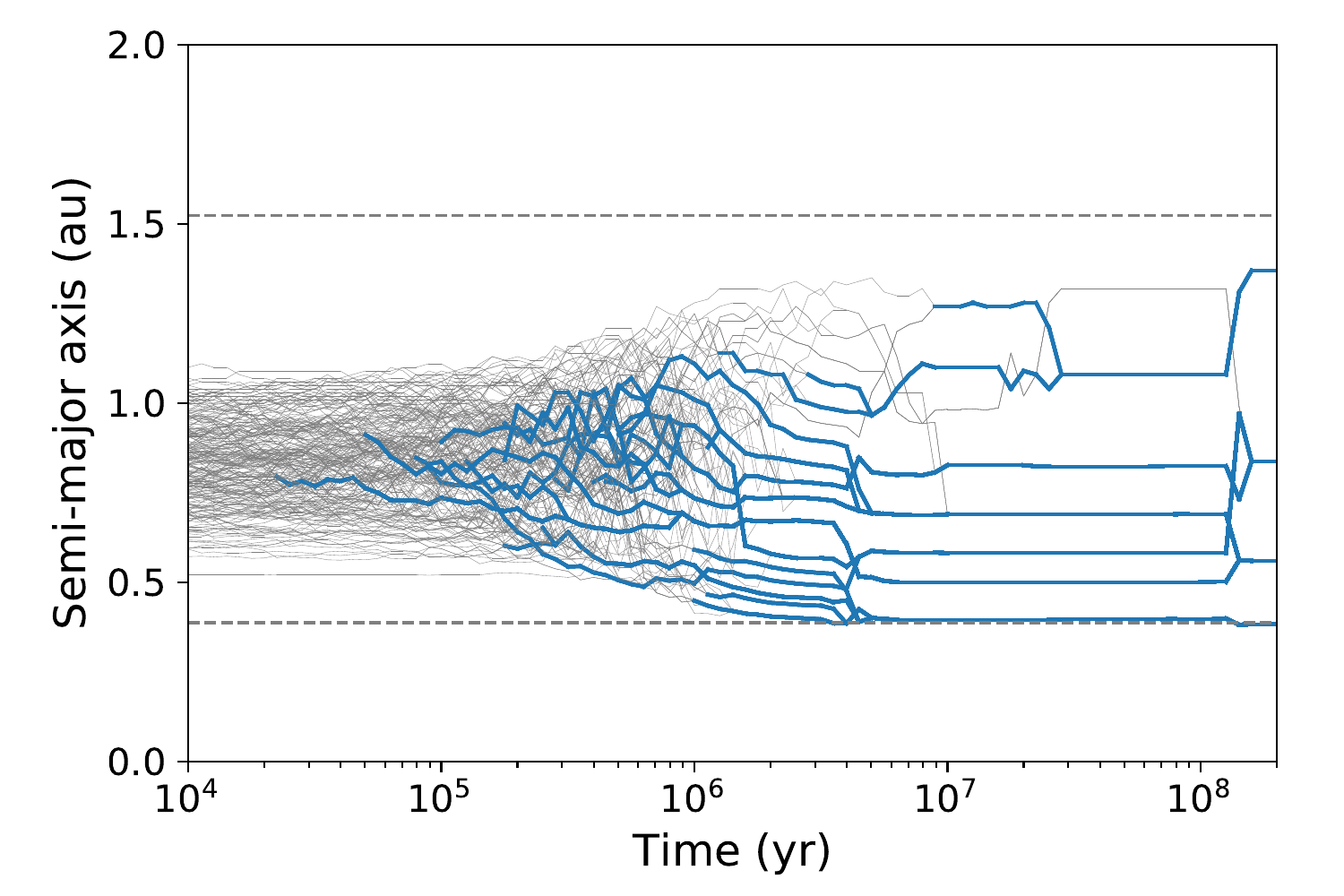}
\caption{
Time evolution of semi-major axis. 
Dashed lines indicate semi-major axes of Mercury and Mars.
Planets with masses greater than Mercury mass are indicated by thick lines.
}
\label{fig:nbody}
\end{center}
\end{figure}

Figure \ref{fig:nbody} shows the time evolution of the semi-major axes of the protoplanets.
The protoplanets collide with each other and eventually form four planets with masses greater than the Mercury mass at 200 Myr after the formation of protoplanets.
Although the turbulent viscosity dominates the early disk evolution, the magnetically-driven disk winds clear the disk gas efficiently in the later phase ($\gtrsim10^{6}$ yr).
Owing to the inner-disk-clearing by the magnetically-driven disk winds, the inner disk becomes flatter and finally dissipates within $10^{7}$ yr (Figure \ref{fig:disk}). 
This flat (or even positive) density profile induces a positive co-rotation torque on the protoplanets, and hence, even massive protoplanets of masses similar to Earth mass do not significantly migrate (see Appendix~\ref{sec:app}).

\begin{figure}[ht]
\begin{center}
\includegraphics[scale=0.3]{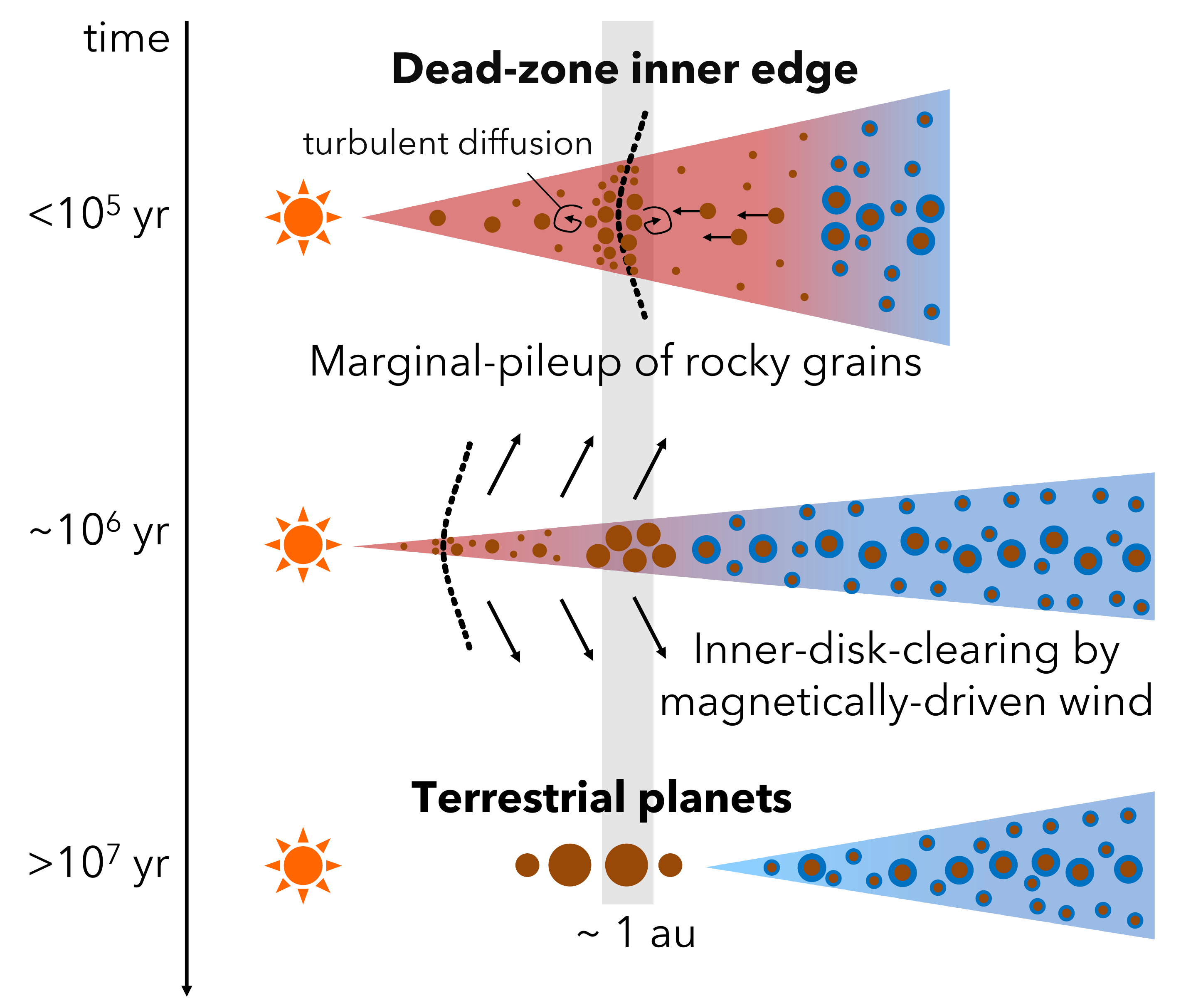}
\caption{
Schematic of terrestrial planet formation starting at dead-zone inner edge.
}
\label{fig:schematic}
\end{center}
\end{figure}
Here we summarize our model in Figure \ref{fig:schematic}.
In the early phase of the disk evolution, the dead-zone inner edge is located beyond 1 au owing to the efficient accretion heating.
Rocky planetesimals form in the terrestrial planet region by the dust pileup at the dead-zone inner edge within 0.1 Myr.
The total mass of the rocky planetesimals is sensitive to the turbulence strength.
When $\alpha_{\rm DEAD}=2.3\times10^{-4}$, rocky planetesimals with a total mass of $\sim 2M_{\oplus}$ form in a narrow annulus at 1 au, which is preferable for the formation of inner solar system planets.
The formed planetesimals can grow into planets without significant migration because the inner disk rapidly dissipate by the magnetically-driven disk winds.

\section{Discussions}\label{sec:Discussion}

\subsection{Features of inner solar system}
\begin{figure}[ht]
\begin{center}
\includegraphics[scale=0.55]{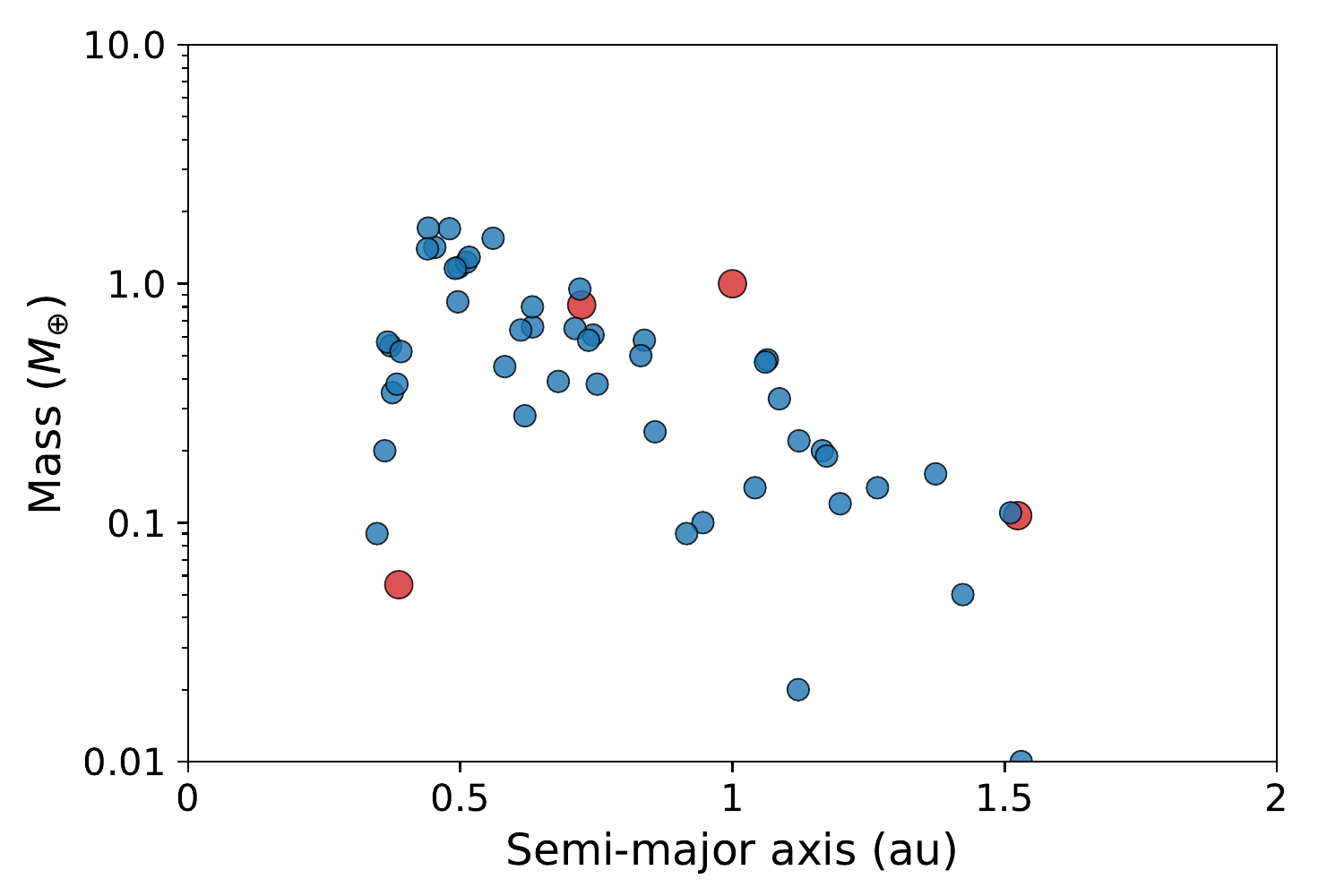}
\caption{
Comparison of mass distributions obtained from ten runs of our simulations (blue) and solar system terrestrial planets (red).
}
\label{fig:stat}
\end{center}
\end{figure}

It is noteworthy that our model can reproduce various characteristics of the inner solar system. Figure \ref{fig:stat} summarizes ten runs of \textit{N}-body simulations that start with different initial locations of the protoplanets.
The important feature that no planets form within the orbit of Mercury was reproduced. 
Regarding the mass distribution, the innermost and outermost planets are less massive, which is consistent with the low masses of Mercury and Mars.

Our model also has important implications for planetary composition. 
The water snow line can be located within 1 au during disk evolution \citep{Oka+11}. 
If terrestrial planets form during this phase, they might be enriched by water. 
To prevent icy materials (i.e., pebbles and planetesimals) from growing near 1 au, the gas disk around the snow line should be dry \citep{Morbidelli+16}. 
However, our model proposes that rocky planetesimals form before the snow line moves in; therefore, there is no need to assume a dry disk gas during the formation of planetesimals. 

As a different problem, there is a possibility that icy pebbles form in the outer region and experience a radial drift to $\sim$ 1 au, making the terrestrial planets water-rich. The radial drift of icy pebbles can be hindered by the formation of a proto-Jupiter \citep{Morbidelli+16,Kruijer+17}.
A potential mechanism forming a Jupiter core is icy planetesimal formation immediately behind the water snow line (e.g., \citealt{DA17,DD18,Hyodo+19}).
Therefore, we expect that a proto-Jupiter was formed at the water snow line immediately after the terrestrial planets form at the dead-zone inner edge.

Other features of the terrestrial planets in the solar system can also be reproduced. The angular momentum deficit \citep{Laskar1997}, which represents the magnitude of the eccentricity and inclination, is as small as $\sim 10^{-3}$, which is consistent with that of the current inner solar system. 
In addition, last giant impacts experienced by massive terrestrial planets with masses of $\simeq 1 \,M_\oplus$ are also consistent with the timing of the moon-forming impact of 50--150 Myr \citep[e.g.,][]{Kleine+09}.

\subsection{Importance of early phase disk evolution}

We demonstrated that inner solar system analogues can form under a specific disk condition.
As shown in Section \ref{sec:param}, the planetesimal distribution is sensitive to disk parameters.
Particularly, the total mass of planetesimals is very sensitive to the turbulence strength in the dead-zone.
Although we fix $\alpha_{\rm DEAD}$, it would vary with time because of fluctuations in the gas motion and of different mechanisms that generate the turbulence.
The small variation in the turbulence strength might affect the total mass of planetesimals.
To fully understand the formation of planetesimals via the dust-pileup, detailed (magneto-)hydrodynamical simulations combined with the dust evolution would be necessarily.

The initial disk size is also important particularly for the radial position of planets.
The initial disk parameters strongly depend on the disk formation processes and parent cloud properties, such as an angular momentum of the collapsing core (e.g., \citealt{HG05}).
To form a compact disk with a radius of $\sim$ 15 au, the angular velocity of the molecular cloud core needs to be $\sim5\times10^{-15}~{\rm s^{-1}}$ which is comparable to the typical observed value \citep{Goodman+93}.
In this study, we do not consider the disk formation phase and assume a power-law disk as an initial condition. 
However, the dust grains grow and potentially form planetesimals even in the disk formation phase \citep{DD18}.
The detailed parameter study considering the disk formation phase would be important for understanding the diversity of exoplanets and will be performed in near future.

The evolution of the global magnetic field is also crucial for both disk formation and evolution.
Our model require inner-disk-clearing by the magnetically-driven disk winds, which depend on the strength of the large-scale magnetic field threading the disk (e.g., \citealt{Bai16}).
In addition, the wind-driven accretion can suppress the mid-plane heating because heat energy is released at the upper layer \citep{Mori+19}.
However, we expect that the location of the dead-zone inner edge would not be affected by the non-ideal magneto-hydrodynamical effect because it takes place only outside the dead-zone inner edge.

It is worth to note that recent ALMA observations have shown that young protoplanetary disks have substructures in dust emission (e.g., \citealt{SE18,Segura-Cox+20,Nakatani+20}).
The prevalence of substructures suggests that local dust trapping occurs and planetesimals might form within them in the early phase of the disk evolution.
Although most of the observed substructures are in the outer region ($\gtrsim 10$ au), it is still unclear if the inner disk also has substructures.
Observations of the terrestrial planet regions are still difficult with current observing facilities, such as ALMA, because we require $<0.01$ arcsec resolution and observing wavelengths where the disk is optically thin.
Future sub-cm observations with such as the next generation Very Large Array (ngVLA) will provide insights on terrestrial planets formation.

\subsection{Comparison to other models with dead-zone inner edge}
In this work we demonstrated that the inner solar system analogues form at the dead zone inner edge if the disk is initially compact and the turbulence in the dead-zone can keep balance between dust accumulation and diffusion.
In contrast, most of previous studies focus on the dead-zone inner edge as a formation site of close-in planets (e.g., \citealt{CT14,Hu+18,Jankovic+19}). 
One of the biggest difference between ours and the previous models is the disk surface density.
In our model we consider a compact and massive disk where the gas surface density is high enough to heat the mid-plane at 1 au above $>$ 800 K, while previous studies have used less massive disks.
It would be natural to focus on the compact disk because dust at 1 au is expected to grow within a timescale of $\sim$ 100 yr and protoplanetary disks are expected to be more compact in earlier phase.

Furthermore, the previous studies mostly consider the regime where dust particles are easily trapped at the dead-zone inner edge. 
However, dust particles are not necessarily trapped even if the radial gas pressure profile has a pressure maximum because turbulent diffusion prevent it \citep{Ueda2019}.
Our work showed that the marginal dust-trapping at the dead-zone inner edge account for the inner solar system, not close-in super earths.
Since the inner solar system analogues form only with a tight parameter space, they might be rare compared to super-earth systems.

\section{Conclusions}\label{sec:Conclusion}

In the early stage of disk evolution, owing to the high surface density, the gas disk accretion heats the disk mid-plane, and the dead-zone inner edge is located beyond 1 au.
As the disk evolves, the surface density decreases and the dead-zone inner edge moves inward.
With the inward migration of the dead-zone inner edge, dust particles pile up at the dead-zone inner edge and planetesimals form.
We found that under specific conditions --- initial disk mass of $0.1M_{\odot}$, initial disk radius of $15~{\rm au}$, and turbulence strength in the dead zone of $2.3\times10^{-4}$ --- rocky planetesimals with a total mass of $\sim$ 2 $M_{\oplus}$ can form in the current terrestrial planet region, with no planetesimals inside $r \simeq 0.6\,{\rm au}$ within 0.1 Myr.
There are two regimes in planetesimal formation at the dead-zone inner edge: no rocky planetesimals form due to efficient turbulence mixing ($\alpha_{\rm DEAD}>2.3\times10^{-4}$) and significant amout of rocky planetesimals form by efficient dust trapping ($\alpha_{\rm DEAD}<2.3\times10^{-4}$).
The inner solar system lies in between these two regimes.
We also found that protoplanets can grow into planets without significant migration in the disk structure because the magnetically-driven disk winds rapidly dissipate the inner disk gas and suppress Type-I migration.
Based on these results, we propose a hypothesis that the solar system terrestrial planets formed by dust trapping at the dead-zone inner edge.
This model can explain the origins of various features of the terrestrial planets in the solar system including the mass and semi-major axis distribution and the rocky composition.

\section*{Acknowledgements}
We thank Masanobu Kunitomo for useful comments.
T.U. is supported by JSPS KAKENHI Grant Numbers JP19J01929.
M.O. is supported by JSPS KAKENHI Grant Numbers 18K13608 and 19H05087.
E. K. and S. O. are supported by JSPS KAKENHI Grant Number 18H05438.
Numerical computations were in part carried out on PC cluster at Center for Computational Astrophysics of the National Astronomical Observatory of Japan.

\appendix

\section{Effect of co-rotation torque on suppression of migration}\label{sec:app}

\begin{figure}[ht]
\begin{center}
\includegraphics[scale=0.55]{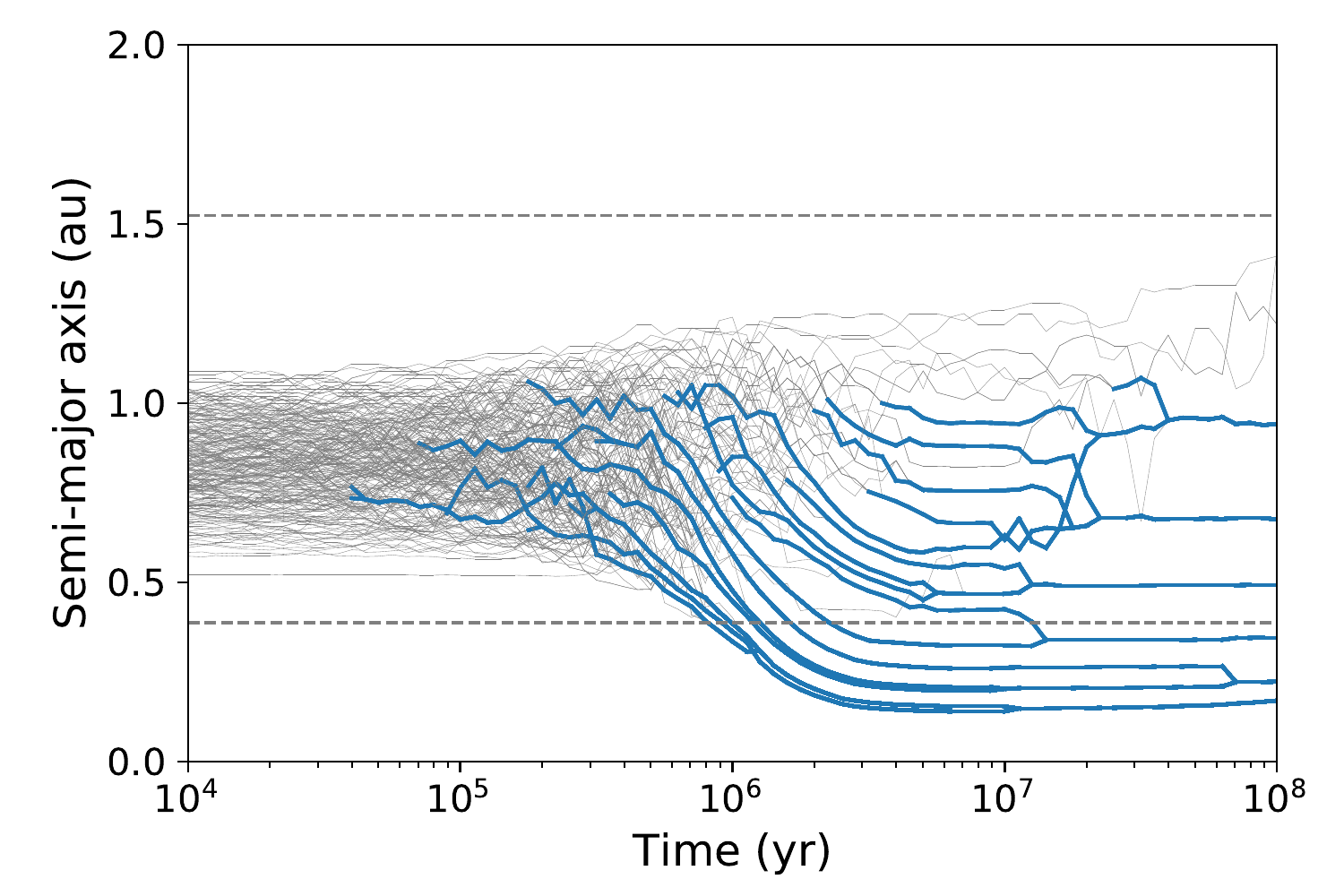}
\caption{Same as Figure\,\ref{fig:nbody} but without effects of co-rotation torque.
}
\label{fig:nbody3}
\end{center}
\end{figure}

It is known that planets with masses larger than the Mars mass undergo rapid inward migration. Therefore, even if protoplanets form in a narrow annulus at $\sim$ 1\,au, they are cuased to penetrate the inner region by the subsequent migration. Note that the disk-planet interaction was not considered in \citet{Hansen2009}.
In Section~\ref{sec:plts2planets-results}, we showed that Type-I migration of planets is significantly suppressed and no planets migrate inside the orbit of Mercury. This is partly because the gas surface density in the inner region is reduced by the disk winds. In addition, the slope of the gas surface density is changed by the disk winds, which also plays an important role in suppressing the inward migration. Here, we see the effect of the positive co-rotation torque due to the change in the gas surface density profiles.
Figure~\ref{fig:nbody3} shows the result of a simulation in which the effect of the co-rotation torque is artificially ignored. 
Comparing the results of the simulations shown in Figures~\ref{fig:nbody} and \ref{fig:nbody3}, an inward migration of the planets are clearly visible in Figure \ref{fig:nbody3}. 
Consequently, the planets move inside the orbit of Mercury. This result confirms that the change in the surface density profile is important for the suppression of Type-I migration.

\bibliographystyle{./aasjournal}
\bibliography{./reference}

\end{document}